# Functional analysis via extensions of the band depth

## Sara López-Pintado[1] and Rebecka Jornsten[2],*

*Universidad Pablo de Olavide and Rutgers University*

**Abstract:** The notion of data depth has long been in use to obtain robust location and scale estimates in a multivariate setting. The depth of an observation is a measure of its centrality, with respect to a data set or a distribution. The data depths of a set of multivariate observations translates to a center-outward ordering of the data. Thus, data depth provides a generalization of the median to a multivariate setting (the deepest observation), and can also be used to screen for extreme observations or outliers (the observations with low data depth). Data depth has been used in the development of a wide range of robust and non-parametric methods for multivariate data, such as non-parametric tests of location and scale [Li and Liu (2004)], multivariate rank-tests [Liu and Singh (1993)], non-parametric classification and clustering [Jornsten (2004)], and robust regression [Rousseeuw and Hubert (1999)].

Many different notions of data depth have been developed for multivariate data. In contrast, data depth measures for *functional data* have only recently been proposed [Fraiman and Muniz (1999), López-Pintado and Romo (2006a)]. While the definitions of both of these data depth measures are motivated by the functional aspect of the data, the measures themselves are in fact invariant with respect to permutations of the domain (i.e. the compact interval on which the functions are defined). Thus, these measures are equally applicable to multivariate data where there is no explicit ordering of the data dimensions. In this paper we explore some extensions of functional data depths, so as to take the ordering of the data dimensions into account.

## 1. Introduction

In functional data analysis, each observation is a real function $x_i$, $i = 1, \ldots, n$, defined on a common interval in $\mathbf{R}$. Functional data is observed in many disciplines, such as medicine (e.g. EEG traces), biology (e.g. gene expression time course data), economics and engineering (e.g., financial trends, chemical processes). Many multivariate methods (e.g. analysis of variance, and classification) have been extended to functional data (see Ramsay and Silverman [22]). A basic building block of such statistical analyses is a location estimate, i.e. the mean curve for a group of data objects, or data objects within a class. When analyzing functional data, outliers can affect the location estimates in many different ways, e.g. altering the shape and/or magnitude of the mean curve. Since measurements are frequently noisy, statistical analysis may thus be much improved by the use of *robust* location estimates, such as

*Corresponding author is funded by NSF Grant DMS-0306360 and EPA Star Grant RD-83272101-0.

[1]Departamento de Economia, Metodos Cuantitativos e Historia economica, Universidad Pablo de Olavide, Edif. n3-Jose Monino-3 planta Ctra de Utrera, Km. 1 41013 Sevilla, Spain, e-mail: sloppin@upo.es

[2]Department of Statistics, Rutgers University, Piscataway, NJ 07030, USA, e-mail: rebecka@stat.rutgers.edu

the median or trimmed mean curve. Data depth provides the tools for constructing these robust estimates.

We first review the concept of data depth in the multivariate setting, where data depth was introduced to generalize order statistics, e.g. the median, to higher dimensions. Given a distribution function $F$ in $\mathbf{R}^d$, a statistical depth assigns to each point $x$ a real, non-negative bounded value $D(x|F)$, which measures the centrality of $x$ with respect to the distribution $F$. Given a sample of n observations $X = \{x_1, \ldots, x_n\}$, we denote the sample version by $D(x|F_n)$ or $D(x|X)$. $D(x|X)$ is a measure of the centrality of a point $x$ with respect to the sample $X$ (or the empirical distribution function $F_n$). The point $x$ can be a sample observation, or constitute independent "test data". For $x = x_i \in X$, $D(x_i|F_n)$ provides a center-outward ordering of the sample observations $x_1, \ldots, x_n$.

Many depth definitions have been proposed for multivariate data (e.g. Mahalanobis [19], Tukey [26], Oja [20], Liu [12], Singh [25], Fraiman and Meloche [3], Vardi and Zhang [27] and Zuo [31]). To illustrate the data depth principle and the variety of depth measures, we will briefly review two very different notions of depth: the simplicial depth of Liu [12], and the $L_1$ depth of Vardi and Zhang [27] (a detailed discussion of the different types of data depths can be found in Liu, Parelius and Singh [14] and Zuo and Serfling [32]). To compute the simplicial depth of a point $x \in \mathbf{R}^d$ with respect to the sample $X = \{x_1, \ldots, x_n\}$, we start by partitioning the sample into a set of $\binom{n}{d+1}$ unique (d+1)-simplices. Consider the two-dimensional case illustrated in Figure 1a. We depict a subset of the $\binom{n}{3}$ 3-simplices (triangles) in $\mathbf{R}^2$ defined by a set of objects $(x_1, x_2, x_3) \in X$. A point $x$ is considered deep within the sample $X$ if many simplices contain it, and vice versa. Formally, the simplicial depth of a point $x$ is defined as

$$SD(x|X) = \binom{n}{d+1}^{-1} \sum_{1 \leq i_1 < i_2 < \cdots < i_{d+1} \leq n} I\{x \subset simplex(x_{i_1}, \ldots, x_{i_{d+1}})\},$$

where $I\{A\}$ is an indicator of the event $A$, equal to 1 if $A$ is true and 0 otherwise. We can see from Figure 1a that the point marked with a triangle is covered by many simplices, resulting in a high depth measure, whereas the point marked with a plus attains the minimum depth measure (i.e. $\binom{n}{3}^{-1}$ if the point is a sample observation, and 0 otherwise).

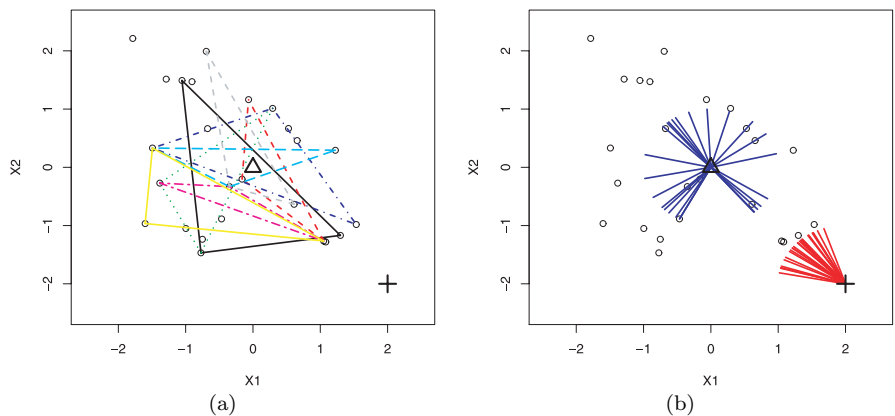

(a)                                    (b)

Fig 1. *(a) The simplicial data depth, (b) The $L_1$ data depth.*

To calculate the $L_1$ data depth of a point $x$ with respect to the sample $X$, we start by forming the unit vectors $e(x, x_i)$ that point from $x$ to $x_i \in X$ (Figure 1b). The $L_1$ depth of $x$ is defined as

$$LD(x|X) = 1 - \bar{e}(x), \quad \text{where} \quad \bar{e}(x) = \frac{1}{n} \sum_{x_i \in X} e(x, x_i).$$

I.e., $\bar{e}(x)$ is the sample average of the unit vectors $e(x, x_i)$. If $x$ is on the periphery of the sample, all $e(x, x_i)$ unit vectors point in an almost identical direction such that $\bar{e}(x) \simeq 1$, and $LD(x|X) \simeq 1 - 1 = 0$ (point marked with a plus in Figure 1b). If $x$ is in the center of the sample, the unit vectors $e(x, x_i)$ will point in many different directions and almost cancel out in the computation of $\bar{e}(x)$, resulting in a high depth measure $LD(x|X) \simeq 1 - 0 = 1$ (point marked with a triangle in Figure 1b).

Focusing on the case when $x$ is a sample observation, $x \in$ X, we see from the above examples that data depth can be used to rank-order the data set $X$, from the deepest to the least deep. We classify $x$ with high $D(x|X)$ as the most representative of the sample $X$, and $x$ with low $D(x|X)$ as the most extreme observations, that may be considered outliers. The deepest observation is a generalization of the median to a multivariate setting, and the center-outward ordering can be used to construct trimmed mean estimates. Robust multivariate estimates based on data depth have been used in a wide range of non-parametric analyses, such as non-parametric testing of location and/or scale (Li and Liu [11]), multivariate rank-test (Liu and Singh [15]), non-parametric classification and clustering (Jornsten [10]), and robust regression (Rousseeuw and Hubert [23]).

In this paper we discuss data depth measures for functional data. We review the *band depths* of López-Pintado and Romo [16] (Section 2), and propose some extensions of these depths to better address the functional characteristic of the data (section 3). We also propose a re-sampling based estimation scheme that speeds up the data depth calculation for sample objects, which is otherwise a computationally intensive task for large data sets (Section 4). In Section 5 we compare the performance of the new notions of depth to that of the band depth, under various simulation scenarios. While we cannot identify a depth measure that dominates all other measures across all simulation scenarios, we do find that data depths that account for the functional characteristics of the data can improve on data depths that do not have this property.

## 2. The band depth and the generalized band depth

In recent years, some definitions of depth for functional data have been proposed. Fraiman and Muniz [3] considered a concept of depth based on the integral of univariate depths. López-Pintado and Romo [16] introduced the notion of *band depth*, which is based on the graphs of the curves and the bands they delimit in the plane. Let $X = \{x_1, \ldots, x_n\}$ be a sample of continuous curves defined in the compact interval $T$. The graph of a function $x$ is given by

$$G(x) = \{(t, x(t)) : t \in T\},$$

where $x$ is either an observation from the sample, or independent test data.

The *band* in $\mathbf{R}^2$ is the region delimited by $j$ curves $x_{i_1}, \ldots, x_{i_j}$, and defined as

$$B(x_{i_1}, x_{i_2}, \ldots, x_{i_j}) = \left\{ (t, y) : \ t \in T, \ \min_{r=1, \ldots, j} x_{i_r}(t) \leq y \leq \max_{r=1, \ldots, j} x_{i_r}(t) \right\} =$$

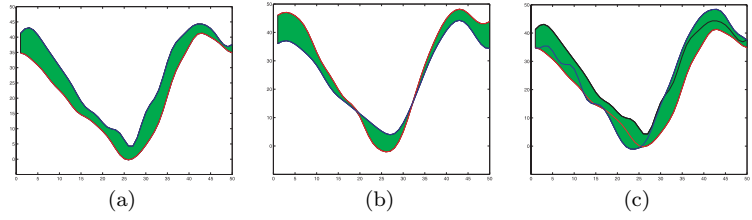

FIG 2. *(a) Band determined by two curves, (b) Band determined by two curves, where the curves cross, and (c) Band determined by three curves.*

$$= \left\{ (t, y) : \ t \in T, \ y = \alpha_t \min_{r=1,\dots,j} x_{i_r}(t) + (1 - \alpha_t) \max_{r=1,\dots,j} x_{i_r}(t), \alpha_t \in [0, 1] \right\}.$$

Here, the $j$ curves $x_{i_1}, \dots, x_{i_j}$ are chosen from the sample $x_1, \dots, x_n$. A sample of size $n$ can thus generate $\binom{n}{j}$ possible bands. Figures 2a and 2b show examples of bands determined by two curves. In Figure 2a the curves form a band that has a non-zero width across the entire compact interval $T$. In Figure 2b the curves cross, and the band is degenerate (has width 0) at the points of crossing. In Figure 2c we depict a band determined by three curves.

**Definition 1.** Given a sample of curves $x_1, \dots, x_n$, the *band depth* $(BD)$ of any curve $x$ is

$$(2.1) \qquad S_{n,J}(x|X) = \sum_{j=2}^{J} S_n^{(j)}(x|X), J \geq 2,$$

where

$$(2.2) \quad S_n^{(j)}(x|X) = \binom{n}{j}^{-1} \sum_{1 \leq i_1 < i_2 < \dots < i_j \leq n} I\{G(x) \subset B(x_{i_1}, x_{i_2}, \dots, x_{i_j})\}, \ j \geq 2.$$

That is, the band depth of object $x$ is defined as the proportion of bands delimited by $j$ curves $(B(x_{i_1}, x_{i_2}, \dots, x_{i_j}))$ containing the graph of $x$. In a multivariate setting, the band depth resembles the simplicial depth of Liu [12]. In fact, in $\mathbf{R}^2$ the band $B(x_{i_1}, x_{i_2}, \dots, x_{i_j})$ corresponds to the smallest *rectangle* with sides parallel to the axes that contain $j$ objects $x_{i_1}, \dots, x_{i_j}$, compared with a triangle (3-simplex) used in the definition of the simplicial depth.

In a functional data setting, the intuition behind the band depth is as follows: If a curve $x$ has a shape that differs from sample curves $x_i \in X$, then few bands can contain it, and vice versa. Thus, a curve that is not representative of the sample will be associated with a low band depth value, and a representative curve will be associated with a high band depth value. The median curve defined by the band depth is $x_{deepest} = argmax \ S_{n,J}(x_i|X)$.

For $J = 2$ the band depth is simple and fast to compute. However, there are some practical limitations: The curves $x_{i_1}, x_{i_2}$ that determine the band $B(x_{i_1}, x_{i_2})$ may cross (Figure 2b). At the point of crossing $t_c$, the band $B(x_{i_1}, x_{i_2})$ is degenerate. Unless a curve $x$ coincides with $x_{i_1}$ and $x_{i_2}$ at the cross-point $t_c$, curve $x$ is not contained in the band. If many curves in the sample cross, most bands $B(x_{i_1}, x_{i_2})$ will not contribute to the band depth measure (equation 2.2). This generates many ties between data objects, and may thus result in a non-unique median curve and center-outward rank-order of the data objects. This limits the practical use of the band depth to construct robust estimates for functional analysis, e.g. non-parametric testing or classification.

If we use $J = 3$, i.e. let 3 curves define the delimiting band $B(x_{i_1}, x_{i_2}, x_{i_3})$, we reduce the impact of curves that cross, and reduce the number of ties. However, the band depth with $J > 2$ is computationally intensive to work with since the number of unique bands grows at rate $\binom{n}{J}$. In addition, the band delimited by 3 data objects (Figure 2c) does not provide the same intuition as the band delimited by 2 objects (Figure 2a). The shape of the band delimited by $J = 3$ objects may differ substantially from the individual objects. Thus, if curve $x$ is contained in band $B(x_{i_1}, x_{i_2}, x_{i_3})$, this does not necessarily imply that $x$ is similar to any of the objects $x_{i_1}, x_{i_2}, x_{i_3}$.

When the curves are very irregular, the band depth can be too restrictive. Few bands will fully contain a data object. This will again result in too many ties between data objects, and a poorly defined center-outward ranking. A more flexible notion of depth (the *generalized* band depth) was therefore proposed in López-Pintado and Romo [16]. It is obtained by replacing the indicator function in definition (2.2) by the *proportion* of time the curve $x$ is inside a corresponding band. Given the sample $x_1, \ldots, x_n$, let for any curve $x$

$$A(x; x_{i_1}, x_{i_2}, \ldots, x_{i_j}) = \left\{ t \in T : \min_{r = i_1, \ldots, i_j} x_r(t) \leq x(t) \leq \max_{r = i_1, \ldots, i_j} x_r(t) \right\}, \; j \geq 2,$$

be the set of points in the interval $T$ where the function $x$ is inside the band delimited by $x_{i_1}, x_{i_2}, \ldots, x_{i_j}$. If $\lambda$ is the Lebesgue measure in $\mathbf{R}$, $\frac{\lambda(A_j(x))}{\lambda(T)}$ is the proportion of time that $x$ is inside the band. We define

$$(2.3) \qquad GS_n^{(j)}(x|X) = \binom{n}{j}^{-1} \sum_{1 \leq i_1 < i_2 < \cdots < i_j \leq n} \frac{\lambda(A(x; x_{i_1}, x_{i_2}, \ldots, x_{i_j}))}{\lambda(T)}, \; j \geq 2,$$

as a generalized version of $S_n^{(j)}(x|x_1, \ldots, x_n)$.

**Definition 2.** Given a sample of curves $x_1, \ldots, x_n$, the *generalized band depth* (*GBD*) of a curve $x$ is

$$(2.4) \qquad\qquad GS_{n,J}(x|X) = \sum_{j=2}^{J} GS_n^{(j)}(x|X), \; J \geq 2.$$

A curve $x$ may be representative of the sample $X$, with the exception of an isolated set of points $t \in T$. The generalized band depth assigns a high depth measure to such observations, whereas the band depth ascribes a minimum depth value. In addition, for $J = 2$ and irregular data, the delimiting objects of a band may cross. While for the band depth, such bands do not contribute to the computation of the depth measures, the generalized band depth will simply down-weigh the contribution of those bands by the proportion of cross-points $t_c$ in $T$. The *GBD* thus computes the data depth by considering each data dimension $t$ separately, similar to the data depth proposed by Fraiman and Muniz [3].

## 3. Extensions of the band depth

The definitions of the band depth and the generalized band depth allow for delimiting objects of a band to cross. We saw that this poses a problem when applying the band depth to noisy and irregular curve data, creating too many ties between the sample curves. Moreover, by allowing the delimiting objects to cross, we lose some of the intuition behind the band depth: The delimiting objects should gather curves of similar shape within the band, such that curves that are contained in the most bands are the most representative of the sample. If delimiting objects are allowed to cross, the shape of the band may in fact differ substantially from the shapes of the sample objects.

In addition, while the generalized band depth allows for excursions outside a band, it makes no distinction between randomly scattered excursions and excursions over a contiguous region. Clearly however, single excursions at random points on the compact interval $T$ are not as informative as an excursion that persists across a set of consecutive points. The latter is more likely to demonstrate that the shape of a curve $x$ differs from those of the band delimiting objects $x_{i_1}, x_{i_2}$.

From the above discussion, and the definitions in Section 2, we see that the band depth ($BD$) and generalized band depth ($GBD$) are both invariant with respect to permutations of the data dimensions, i.e. permutations of $t \in T$. This raises the question whether the performance of these depths can be improved via extensions of the depth measures that take the explicit ordering of the data dimensions for functional data into account. We will explore such extensions of the band depth and generalized band depths in Sections 3.1 and 3.2. In what follows, we focus on the band depth where $J = 2$ delimiting objects form each band, though generalizations to $J > 2$ can be made at the expense of an increased computational burden.

### 3.1. The corrected (generalized) band depth

We begin by revisiting the definition of a delimiting band. Ideally, we want each band to gather curves of similar shape inside it. To achieve this, we make the following adjustments to the definition of a band: If two curves cross, the band will be defined only where one of the curves is the upper curve and the other one is the lower curve. Hence, for each pair of curves that cross, there are two possible bands (depending on which curve is consider the upper curve). We will choose the longest of the two bands. The notions of *corrected* band depth and its generalized version are defined similarly to the band depth and the generalized band depth, but re-defining the band as described above.

We first define the corrected band depth ($cBD$). Let $a(i_1, i_2) = \{t : x_{i_2} - x_{i_1} \geq 0\}$ and $L_{i_1, i_2} = \frac{\lambda(a(i_1, i_2))}{\lambda(T)}$ ($\lambda$ is the Lebesgue measure). By exchanging the roles of the upper ($x_{i_1}$) and lower ($x_{i_2}$) curves we obtain $L_{i_2, i_1}$ in a similar fashion. We define the corrected band $B_c$ as

$$B_c = I_{\{L_{i_1, i_2} \geq 1/2\}} B_c(x_{i_1}, x_{i_2}) + I_{\{L_{i_2, i_1} > 1/2\}} B_c(x_{i_2}, x_{i_1}),$$

where

$$B_c(x_{i_1}, x_{i_2}) = \{(t, y) : t \in a(i_1, i_2), x_{i_1}(t) \leq y \leq x_{i_2}(t), \},$$

and similarly for $B_c(x_2, x_1)$. We also form the corrected graph $G(x^*)$ as

$$G(x^*) = \{(t, x(t)) : t \in a(i_1, i_2)\}, \quad \text{if} \quad L_{i_1, i_2} \geq 1/2,$$

$$G(x^*) = \{(t, x(t)) : t \in a(i_2, i_1)\}, \quad \text{if} \quad L_{i_2, i_1} > 1/2.$$

We can now define the corrected band depth of a curve $x$ with respect to the sample $X$ as

$$(3.1) \qquad cBD(x|X) = \binom{n}{2}^{-1} \sum_{1 \leq i_1 < i_2 \leq n} \max(L_{i_1,i_2}, L_{i_2,i_1}) I_{\{G(x^*) \in B_c\}}$$

The term $\max(L_{i_1,i_2}, L_{i_2,i_1})$ acts as a *weight*. $cBD$ will thus down-weigh the contribution of bands delimited by curves that cross, but not as drastically as $BD$ where such bands would not contribute at all. In the simulation study (section 4) we see that $cBD$ can substantially improve on the $BD$ when the data is contaminated by curves with different shapes from the rest of the data.

To allow for random excursions of a curve $x$ outside the corrected band $B_c$, as is likely to happen with noisy data and irregular curves, we also propose the corrected generalized band depth as a more flexible alternative. We define the $cGBD$ of a curve $x$ with respect to the sample $X$ as

$$(3.2) \qquad cGBD(x|X) = \binom{n}{2}^{-1} \sum_{1 \leq i_1 < i_2 \leq n} \frac{\lambda(A^c(x; x_{i_1}, x_{i_2}))}{\lambda(T)}.$$

where

$$A^c(x; x_{i_1}, x_{i_2}) = \{t \in a(i_1, i_2) : x_{i_1}(t) \leq x(t) \leq x_{i_2}(t)\}, \quad \text{if } L_{i_1, i_2} \geq 1/2$$
$$= \{t \in a(i_2, i_1) : x_{i_2}(t) \leq x(t) \leq x_{i_1}(t)\}, \quad \text{if } L_{i_2, i_1} > 1/2.$$

Hence, the difference between the *corrected* (generalized) band depth and the (generalized) band depth is that the band is modified in order to consider only the proportion of the domain where the delimiting curves define a contiguous region of non-zero width.

### 3.2. $GBD_I$ and $GBD_O$ — *accounting for consecutive band excursions*

If a curve is only partially contained in a delimiting band, it may still share many characteristics with the delimiting objects. However, this similarity may not be optimally measured by the *number* of excursions outside the band, as with $GBD$, but perhaps how these excursions present themselves. We therefore propose two alternative definitions of the generalized band depth called $GBD_I$ and $GBD_O$.

We propose $GBD_I$ as a more conservative alternative to $GBD$. We replace $\lambda(A(x; x_{i_1}, x_{i_2}))$ (the number of $t \in T$ where $x \in B(x_{i_1}, x_{i_2})$) in equation (2.3) by a measure $CI(x; x_{i_1}, x_{i_2})$, where

$$CI(x; x_{i_1}, x_{i_2}) = \max_{t_S} \left\{ \lambda(t_S) : \min_{r=i_1, i_2} x_r(t) \leq x(t) \leq \max_{r=i_1, i_2} x_r(t), \forall t \in t_S \right\},$$

where $t_S$ is a *compact interval*. That is, $CI(x; x_{i_1}, x_{i_2})$ is the longest consecutive stretch for which $x$ is inside in the band delimited by $x_{i_1}, x_{i_2}$ (<u>C</u>onsecutive <u>I</u>nside). Given sample functions $x_1, x_2, \ldots, x_n$, the $GBD_I$ of a curve $x$ is

$$(3.3) \qquad GBD_I(x|X) = \binom{n}{2}^{-1} \sum_{1 \leq i_1 < i_2 \leq n} \frac{CI(x; x_{i_1}, x_{i_2})}{\lambda(T)}.$$

With $GBD_I$, only the longest non-contaminated portion of a curve $x$ contributes to its depth value. If a curve $x$ weaves in and out of a band, the $GBD$ may still allot a high depth value to $x$, whereas the $GBD_I$ requires that the curve $x$ resides within the band over a large compact set.

We further propose $GBD_O$, where we penalize band excursions that are consecutive. We view a long stretch of a band excursion as evidence that the curve $x$ is not similar to the band delimiting objects $x_{i_1}, x_{i_2}$. Thus, in terms of the total number of band excursions, $GBD_O$ is less conservative than $GBD$. $GBD_O$ penalizes deviations that are persistent, and can thus serve as an indicator that the functional characteristic of curve $x$ differs from the band delimiting objects.

Similar to the $GBD_I$ we start by re-defining the $GBD$ band measure $\lambda(A(x; x_{i_1}, x_{i_2}))$ by an alternative measure $CO(x; x_{i_1}, x_{i_2})$, where

$$CO(x; x_{i_1}, x_{i_2}) = \max_{t_S} \left\{ \lambda(t_S) : \min_{r=i_1, i_2} x_r(t) > x(t) \text{ or } x(t) > \max_{r=i_1, i_2} x_r(t), \forall t \in t_S \right\},$$

and $t_S$ again denotes a compact set on the interval $T$. That is, $CO(x; x_{i_1}, x_{i_2})$ is the longest contiguous region of the curve outside the delimiting band (<u>C</u>onsecutive <u>O</u>utside). The $GBD_O$ of curve $x$ with respect to sample $x_1, \ldots, x_n$ is defined as

$$(3.4) \qquad GBD_O(x|X) = \binom{n}{2}^{-1} \sum_{1 \le i_1 < i_2 \le n} 1 - \frac{CO(x; x_{i_1}, x_{i_2})}{\lambda(T)}.$$

With the depth measure $GBD_O$, a curve $x$ is penalized if the contamination is persistent ($CO(x; x_{i_1}, x_{i_2})$ is large for many bands $B(x_{i_1}, x_{i_2})$), whereas random and isolated excursions are largely ignored. Thus, unlike $GBD$, $GBD_O$ makes a distinction between noise contaminations (random "spikes") and shape contaminations.

## 4. Fast computation via data re-sampling

The computational cost of calculating the data depth of objects in a sample $X = \{x_1, \ldots, x_n\}$ grows with the sample size at rate $\binom{n}{j}$. If the number of curves in the sample of interest is large, the computational burden of data depth based methods puts a serious limit on their applicability. Moreover, non-parametric testing, clustering and classification frequently involve iterative procedures. For example; Li and Liu [11] use bootstrap to compute the sampling distribution of the non-parametric test statistic under the null; the clustering method of Jornsten [10] iterates between updating the cluster center (deepest object) and the cluster allocation, until convergence. For these computationally challenging non-parametric and robust analyses to be competitive with standard approaches, in a practical sense, we need to make each data depth estimation step fast and efficient.

Therefore, we propose a simple method for computing the depth of each curve in a sample based on re-sampling. This re-sampling based data depth calculation is also applicable to multivariate data, and can easily be adapted to other notions of depth.

We begin by dividing the sample $x_1, \ldots, x_n$ into K randomly selected and roughly equal size parts. We refer to each data part as $X_1, \ldots, X_K$, where $\{x_i \in X_k\}$ is a set of $\sim n/K$ objects. For each curve $x$, we obtain independent depth measures with respect to each data part:

$$D(x|X_1), \ldots, D(x|X_K),$$

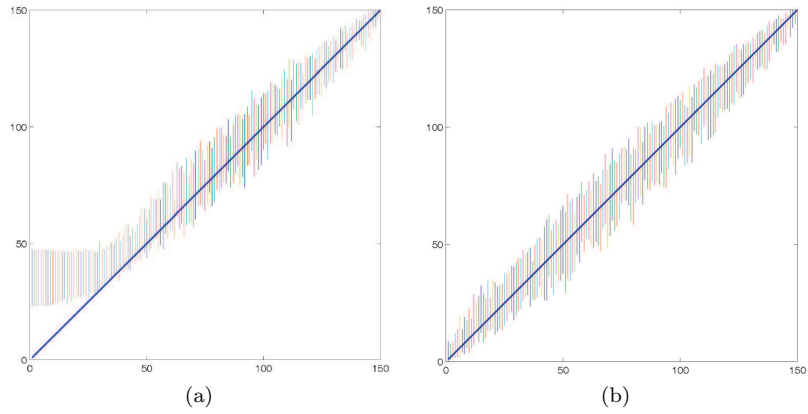

Fig 3. *Re-sampling based rank versus full data rank using (a) BD and (b) cBD.*

where $D$ refers to any of the depth measures discussed in this paper (e.g. $GBD_I$).

We finally define the re-sampling based depth of a curve $x$ as

$$(4.1) \qquad D_r(x|X) = \frac{1}{K} \sum_{k=1}^{K} D(x|X_k)$$

Using simulated data, we investigate the feasibility of replacing the data depth with the re-sampling based version. We generate 150 curves $x_i(t)$ from a model

$$x_i(t) = 4t + e_i(t), \ 1 \le i \le n,$$

where $e_i(t)$ is a sample from a gaussian stochastic process with zero mean and covariance function $\gamma_1(s,t) = \exp\{-|t-s|^2\}$. We compute the depth-based ranks for the sample curves, and the corresponding re-sampling based ranks.

In Figures 3, 4 and 5 we compare the rank-orders induced by the re-sampling based method to the rank-orders obtained with the full data. For each of $b = 1, \ldots, B = 50$ simulated data sets of size $n = 150$, we obtain the ranks of each of the $i = 1, \ldots, n$ data objects from the full data: $D^b(X) = \{D^b(x_1|X), \ldots, D^b(x_n|X)\}$.

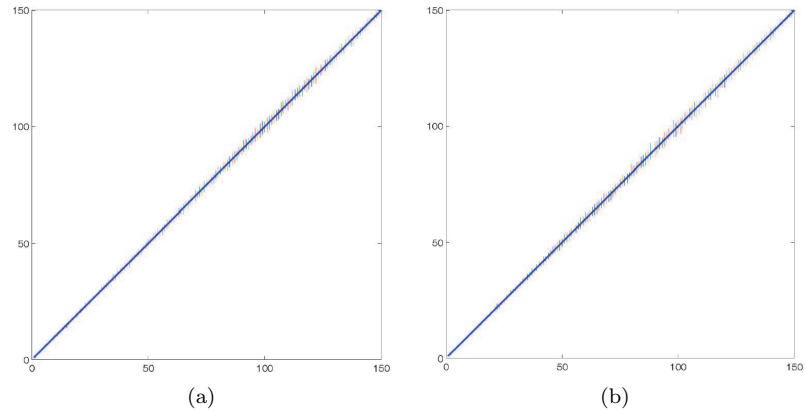

Fig 4. *Re-sampling based rank versus full data rank using (a) GBD, (b) cGBD.*

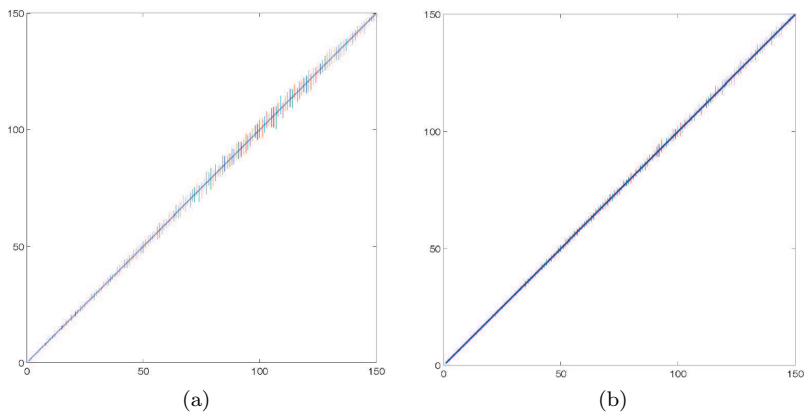

Fig 5. *Re-sampling based rank versus full data rank using (a) $GBD_I$, and (b) $GBD_O$.*

The rank-order is $i_b(1), \ldots, i_b(n)$. We partition the data into $K = 10$ parts, and compute the re-sampling based ranks: $D_r^b(X) = \{D_r^b(x_1|X), \ldots, D_r^b(x_n|X)\}$. We sort the re-sampling based ranks using the full data rank order to obtain: $D_r^{b,*}(X) = \{D_r^b(x_{i_b(1)}|X), \ldots, D_r^b(x_{i_b(n)}|X)\}$. If $D_r^{b,*}(X) \simeq \{1, \ldots, n\}$, the re-sampling based ranks closely agree with the full data ranks. In the figures we plot the mean of the re-sampling based ranks for the 50 simulations against $\{1, \ldots, n\}$. We also depict the standard deviations as vertical bars in the figures. If the re-sampling based depth estimates are competitive with the full data based estimates, we expect the $D_r^{b,*}(X)$ to fall on the line $y = x$ in each plot.

The simulations confirm that, with the exception of the band depth, $BD$, all re-sampling based depth rank-orders agree closely with the full data depth rank-orders. The $BD$ re-sampling based estimates are not accurate for the deepest set of observations since many ties are generated with a smaller sample $\sim n/K$. Using the corrected band depth, $cBD$, alleviates this problem. For the more extreme observations, the $BD$ and $cBD$ rank-orders fall on the $y = x$ line. The $GBD, cGBD, GBD_I$ and $GBD_O$ all show a strong agreement between the full data and re-sampling based depth rank-orders. In addition, the standard deviations are quite small. We obtain similar results in several simulations, including both functional data and multivariate data (omitted to conserve space).

We thus conclude that the ranks obtained from re-sampling based depths are near equivalent to the ranks obtained using the full data to compute the depths. This justifies using the computationally faster re-sampling based versions of the depths in practical applications.

## 5. Simulations and illustrations

In this section we compare our new notions of depth, to the band depth and generalized band depth proposed by López-Pintado and Romo [16], on several simulated data sets. In each simulation, we generate curves from a base model. We then randomly contaminate the data set with six different types of contaminations. The contamination types include those previously analyzed by Fraiman and Muniz [4] and by López-Pintado and Romo [16].

The base model is $X_i(t) = g(t) + e_i(t), 1 \leq i \leq n$, where $e_i(t)$ is a stochastic gaussian process with zero mean and covariance function $\gamma_1(s, t) = \exp\{-|t - s|^{1.5}\}$ and $g(t) = 4t$, with $t \in [0, 1]$. There are many different ways of defining an outlier or a contamination within a sample of curves. For instance, a curve could be very distant from the mean (magnitude outlier) or have a pattern different from the other curves, e.g. decreasing when the remaining ones are increasing, or very irregular in a set of smooth curves (shape outlier). We describe the six different contamination scenarios below (illustrated in Figure 6).

**Model 0.** All curves come from the base model.

**Model 1.** An asymmetric contamination is included in Model 1;

$$Y_i(t) = X_i(t) + c_i M, \ 1 \le i \le n,$$

where $c_i$ is 1 with probability $q$ and 0 with probability $1 - q$; $M$ is the contamination size constant.

**Model 2.** In Model 2 the contamination is symmetric;

$$Y_i(t) = X_i(t) + c_i \sigma_i M, \ 1 \le i \le n,$$

where $c_i$ and $M$ are defined as in Model 1 and $\sigma_i$ is a sequence of random variables independent of $c_i$ taking values 1 and $-1$ with probability $1/2$.

**Model 3.** A partial contamination constitutes Model 3;

$$Y_i(t) = X_i(t) + c_i \sigma_i M, \ \text{if } t \ge T_i, \ 1 \le i \le n,$$

and

$$Y_i(t) = X_i(t), \ \text{if } t < T_i,$$

where $T_i$ is a random number generated from a uniform distribution on $[0, 1]$.

**Model 4.** Model 4 is contaminated by peaks;

$$Y_i(t) = X_i(t) + c_i \sigma_i M, \ \text{if } \ T_i \le t \le T_i + l, \ 1 \le i \le n,$$

and

$$Y_i(t) = X_i(t), \ \text{if } t \notin [T_i, T_i + l],$$

where $l = 2/30$ and $T_i$ is a random number from a uniform distribution in $[0, 1 - l]$. The contamination only occurs for a short subinterval of length $l$.

**Model 5.** We also propose a new model for comparison, where the contamination is like the one in Model 4, but it will occur at $k$ different points uniformly distributed in the domain. Specifically, the model is

$$Y_i(t) = X_i(t) + c_i \sigma_i M, \ \text{if } \ t \in \{T_1, \dots, T_k\},$$

and $Y_i(t) = X_i(t)$ otherwise, where $\{T_1, \dots, T_k\}$ are $k$ random numbers uniformly chosen from the interval $[0, 1]$.

**Model 6.** In Model 6 we consider *shape contaminations* (López-Pintado and Romo [16]). To include shape outliers, we use the covariance function structure proposed in Wood and Chan [29] $\gamma(s, t) = k \exp\{-c|t - s|^\mu\}$, with $s, \ t \in [0, 1]$, and $k$, $c$, $\mu > 0$. Different values of $k$, $c$ and $\mu$ change the shape of the generated functions. For example, increasing $\mu$ and $k$, makes the curves smoother; whereas, increasing $c$ makes the curves more irregular. Model 6 is a mixture of $X_i(t) = g(t) + e_{1i}(t)$, $1 \le i \le n$, with $g(t) = 4t$ and $e_{i1}(t)$ a gaussian stochastic process with zero mean and covariance function $\gamma_1(s,t) = \exp\{-|t-s|^2\}$ and $Y_i(t) = g(t) + e_{2i}(t)$, $1 \leq i \leq n$, with $e_{i2}(t)$ a gaussian process with zero mean and covariance function $\gamma_2(s,t) = \exp\{-|t-s|^{0.2}\}$. The contaminated Model 6 is $Z_i(t) = (1-\varepsilon)X_i(t) + \varepsilon Y_i(t)$, $1 \leq i \leq n$, where $\varepsilon$ is a Bernoulli variable $Be(q)$ and $q$ is a small contamination probability; thus, we contaminate a sample of smooth curves from $X_i(t)$ with more irregular curves from $Y_i(t)$.

We analyze the performance of different notions of depth in terms of robustness. The notions of depth considered are: the band depth with $J = 2$ and $J = 3$ ($BD2, BD3$), the generalized band depth ($GBD$), the corrected version of the band depth and generalized band depth ($cBD, cGBD$), and the generalized band depths that account for consecutive band excursions, $GBD_I$ and $GBD_O$. We compare the mean and the $\alpha$-trimmed mean, given by

$$\widehat{\mu}_n(t) = \frac{\sum_{i=1}^{n} Y_i(t)}{n} \qquad \text{and} \qquad \widehat{m}_n^\alpha(t) = \frac{\sum_{i=1}^{n-[n\alpha]} Y_{(i)}(t)}{n - [n\alpha]},$$

where $Y_{(1)}, Y_{(2)}, \ldots, Y_{(n)}$ is the sample ordered from the deepest to the most extreme (least deep) curve and $[n\alpha]$ is the integer part of $n\alpha$. The rank-orders are computed using the re-sampling based depths. For each model, we consider $R = 200$ replications, each generating $n = 150$ curves, with contamination probability $q = 0.1$ and contamination constant $M = 25$. The integrated errors (evaluated at $V = 30$ equally spaced points in $[0,1]$) for each replication $j$ are

$$EI_\mu(j) = \frac{1}{V}\sum_{k=1}^{V} \left[\widehat{\mu}_n(k/V) - g\left(k/V\right)\right]^2 \text{ and } EI_D^\alpha(j) = \frac{1}{V}\sum_{k=1}^{V} \left[\widehat{m}_n^\alpha(k/V) - g\left(k/V\right)\right]^2,$$

where $D$ refers to one of the data depths ($BD2, BD3, cBD, GBD, cGBD, GBD_I$ or $GBD_O$).

All methods are applied to each simulated data set. Across simulated data sets we see a lot variability since the contaminations are random. We therefore summarize the results as follows; (1) For each modeling scenario and each simulated data set $j$, we compute the minimum integrated error across all methods

$$EI_*^\alpha(j) = \min_{D = BD2, BD3, cBD, GBD, cGBD, GBD_I, GBD_O} EI_D^\alpha(j);$$

(2) We adjust the integrated errors by subtracting the minimum value, such that the best method has adjusted integrated error 0, $EAI_D^\alpha(j) = EI_D^\alpha(j) - EI_*^\alpha(j)$. We summarize the simulation results in terms of the mean and standard deviation of the adjusted integrated errors, $EAI$

We will first examine each simulation model separately, and then discuss the overall results at the close of the section.

Model 0 generates uncontaminated data. From Table 1 we see that the mean is the best estimate, as expected. The generalized band depths ($GBD, cGBD, GBD_I$ and $GBD_O$) perform better than the band depths ($BD2, BD3$ and $cBD$) in this setting, but all of the robust estimates perform reasonably well on uncontaminated data. Some loss of estimation efficiency is unavoidable since a fixed trimming factor $\alpha = 0.2$ was used.

Simulations settings 1 through 3 correspond to simple contaminations, i.e. a positive or negative mean offset, or a partial mean offset contamination. In this

*Simulation results for the seven modeling scenarios. Mean and standard deviations of the adjusted integrated errors (subtracting the integrated error of the winning method for each simulation), with $R = 200$ replications, $q = 0.1$ and $\alpha = 0.2$.*

|          | M0          | M1          | M2          | M3          | M4          | M5          | M6          |
|----------|-------------|-------------|-------------|-------------|-------------|-------------|-------------|
| Mean     | **0.002**   | 6.600       | 0.463       | 0.204       | 0.036       | 0.063       | 0.012       |
|          | **(0.003)** | (3.319)     | (0.606)     | (0.027)     | (0.025)     | (0.025)     | (0.012)     |
| $BD2$    | 0.007       | 4.864       | 0.286       | 0.204       | 0.021       | 0.036       | 0.008       |
|          | (0.009)     | (3.832)     | (0.395)     | (0.278)     | (0.019)     | (0.028)     | (0.010)     |
| $BD3$    | 0.007       | 3.916       | 0.163       | **0.053**   | 0.012       | 0.014       | **0.005**   |
|          | (0.009)     | (3.189)     | (0.227)     | **(0.119)** | (0.014)     | (0.018)     | **(0.005)** |
| $cBD$    | 0.009       | 5.353       | 0.277       | 0.091       | **0.006**   | **0.003**   | 0.005       |
|          | (0.012)     | (4.137)     | (0.365)     | (0.274)     | **(0.011)** | **(0.008)** | (0.010)     |
| $GBD$    | 0.005       | 0.085       | 0.005       | **0.050**   | 0.047       | 0.074       | 0.013       |
|          | (0.005)     | (0.286)     | (0.008)     | **(0.074)** | (0.034)     | (0.034)     | (0.010)     |
| $cGBD$   | 0.004       | 0.117       | 0.005       | **0.048**   | 0.046       | 0.074       | 0.013       |
|          | (0.005)     | (0.381)     | (0.008)     | **(0.075)** | (0.034)     | (0.033)     | (0.010)     |
| $GBD_I$  | 0.005       | 0.260       | 0.007       | **0.055**   | 0.043       | 0.043       | **0.003**   |
|          | (0.005)     | (0.619)     | (0.010)     | **(0.072)** | (0.031)     | (0.026)     | **(0.006)** |
| $GBD_O$  | 0.005       | **0.001**   | **0.003**   | 0.060       | 0.049       | 0.083       | 0.022       |
|          | (0.007)     | **(0.003)** | **(0.007)** | **(0.090)** | (0.036)     | (0.037)     | (0.022)     |

setting we expect only marginal gains from the use of shape sensitive or restrictive depths such as $cBD$ and $GBD_I$. On the other hand, the data is noisy so we expect that $cGBD$ and $GBD_O$ should perform well.

For the case of asymmetric contamination (Model 1), we find that $GBD_O$ outperforms the other methods. The trimming factor is $\alpha = 0.2$, while the contamination probability is $q = 0.1$. Thus, for many of the simulated data sets, some of the uncontaminated curves will be trimmed. The band depths ($BD2, BD3$ and $cBD$) all struggle in this setting. The uncontaminated curves frequently cross, leading to too many ties in the rank-order. While the $GBD, cGBD$ and $GBD_I$ perform much better than the band depths, they are not competitive with $GBD_O$. The source of the problem lies in the asymmetry of the contaminations, and that the magnitude outliers contribute to the computation of the depth values of all other observations (see Figure 6). Since $GBD, cGBD$ and $GBD_I$ are more restrictive than $GBD_O$, these three depth measures trim more curves from the lower portion of the uncontaminated set than from the upper portion, creating a bias in the trimmed mean estimate. This suggests that perhaps an iterative procedure, were outliers are dropped one at a time and the data depths re-estimated after each step, would perform better.

As expected, when the contamination is symmetric (Model 2), $GBD, cGBD$ and $GBD_I$ perform almost as well as $GBD_O$. The trimming of the uncontaminated data set is now mostly symmetric, and the trimmed mean estimates essentially unbiased. The band depths are not competitive.

Model 3 generates data objects that have been partially contaminated. This symmetric contamination is easily identified by $GBD, cGBD, GBD_I$ and $GBD_O$.

Simulation settings 4 through 6 can loosely be seen as shape contaminations (Figure 6). Here, we expect the $cBD$ and $GBD_I$ to perform well, whereas the performances of the less restrictive $GBD, cGBD$ and $GBD_O$ may deteriorate.

Indeed, for Model 4 (peak contamination), $cBD$ is the best method, followed by $BD3$ and $BD2$. $cBD$ improves on the band depths since the band correction accounts for curves that cross. The generalized band depths ($GBD, cGBD, GBD_I$ and $GBD_O$) do not perform well in this setting, even occasionally resulting in an integrated error exceeding that of the mean.

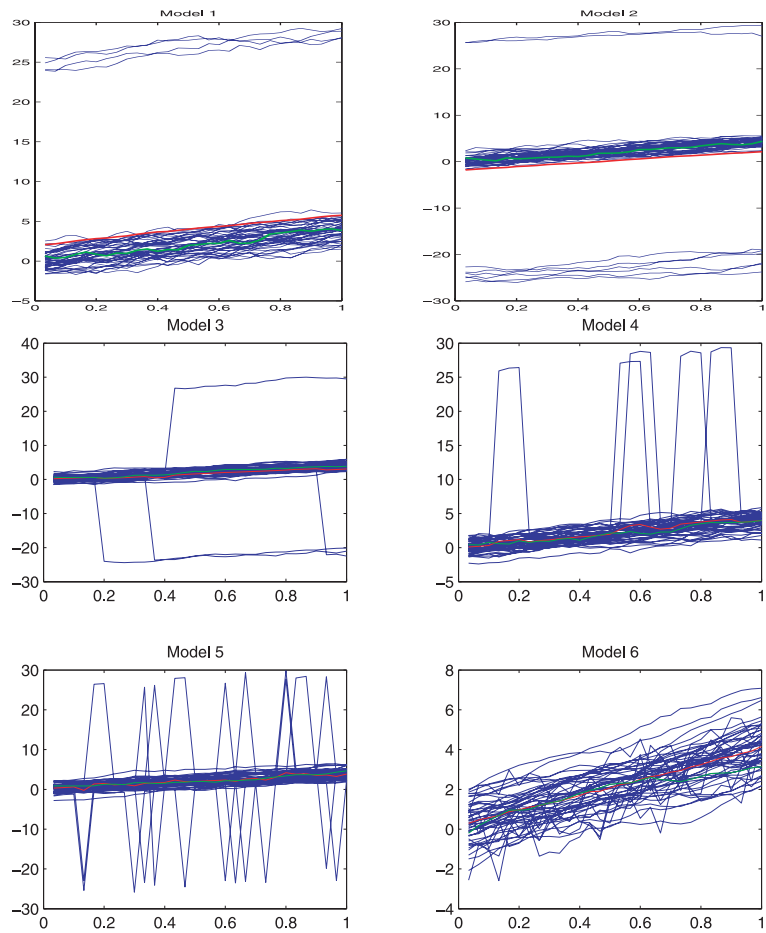

Fig 6. *Curves generated from Model 1: asymmetric contamination (top left), Model 2: symmetric contamination (top right), Model 3: partial contamination (middle left), Model 4: peaks contamination (middle right), Model 5: uniform contamination (bottom left), Model 6: shape contamination (bottom right). The mean curve is depicted in red, and the trimmed mean in green in all cases.*

In Model 5 we allow for multiple contaminations of each curve. The performance of the $GBD, cGBD$ and $GBD_O$ deteriorates since these short excursions are not recognized as a contamination. $GBD_I$ performs a little better, since the uniform contamination leads to the contaminated curves residing in the bands for only short consecutive stretches. Still, $GBD_I$ is not competitive with the band depths (though better than the mean). The best method is again the $cBD$, followed by the computationally expensive $BD3$.

In Model 6 (shape contamination), a set of smooth curves are contaminated by a set of irregular curves. We expect the shape sensitive methods (band depths, $GBD_I$) to excel in this setting. Indeed, the $GBD_I$ is the best method, followed closely by $cBD$ and $BD3$. Again, the less restrictive depths ($GBD, cGBD$ and $GBD_O$) do not perform well in this setting.

In the first three simulations (Models 1 through 3), the generalized band depths, with $GBD_O$ in the lead, outperform the band depths. The contaminations in Models 1 through 3 are essentially magnitude outliers, and persistent across the compact interval $T$. In simulations from Models 4 through 6, the contaminations are more subtle (random peaks, or a different covariance structure). In such settings, the less restrictive data depths, that discard short excursions as non-informative, perform poorly. The band depths, with the corrected band depth in the lead, as well as the $GBD_I$, perform well in this setting.

It is clear then, that one depth cannot be defined as the "best" across all possible scenarios. Therefore, in practise one needs to consider the type of contaminations to screen against. We recommend that several depths are applied to each data set, and the outliers identified by each depth measure examined graphically. From the above simulation results we see that $GBD_O$ and $cBD$ are two candidate measures that are fast and easy to compute, and would highlight different structures in the data.

To illustrate the screening properties of the depth measures, we apply six notions of depth to a data set consisting of the daily temperature in 35 different Canadian weather stations for one year (Ramsay and Silverman [22]). The original data was smoothed using a Fourier basis with sixty five elements. In Figure 7 we show the mean and the median (deepest) curve identified by the $BD, cBD, GBD, cGBD, GBD_I$ and $GBD_O$. In addition, in each figure we highlight the 20% least deep curves in green (wide lines). From the above discussion, and the definitions in Sections 2 and 3, we know that $GBD, cGBD$ and $GBD_O$ are the least restrictive depths. These depths will largely identify magnitude outliers as the least deep. In Figure 7 we see that this is indeed the case for the temperature data set. In contrast, $BD, cBD$ and $GBD_I$ are the most restrictive, and will identify shape outliers. Temperature profiles that are flatter than the rest of the data, or irregular with large, local fluctuations, are identified by these depths.

## 6. Discussion

We introduce several extension of the band depth for functional data. These new notions of depth account for the explicit ordering of the data dimensions inherent to functional data. We find that a simple alteration of the definition of a delimiting band, a band correction, can improve on the previously proposed band depth. In addition, a differential treatment of band excursions that are consecutive versus isolated, can boost the performance of the generalized band depth.

While two of our proposed extensions, the corrected band depth $cBD$ and the $GBD_O$, improve on the band depth and the generalized band depth respectively, we cannot identify a "best" data depth for functional data that uniformly dominates the other measures across all forms of contamination. The $GBD_O$ is the most competitive in simple contamination scenarios (magnitude outliers), whereas the $cBD$ is more competitive when the samples are contaminated by curves of a different structure or shape.

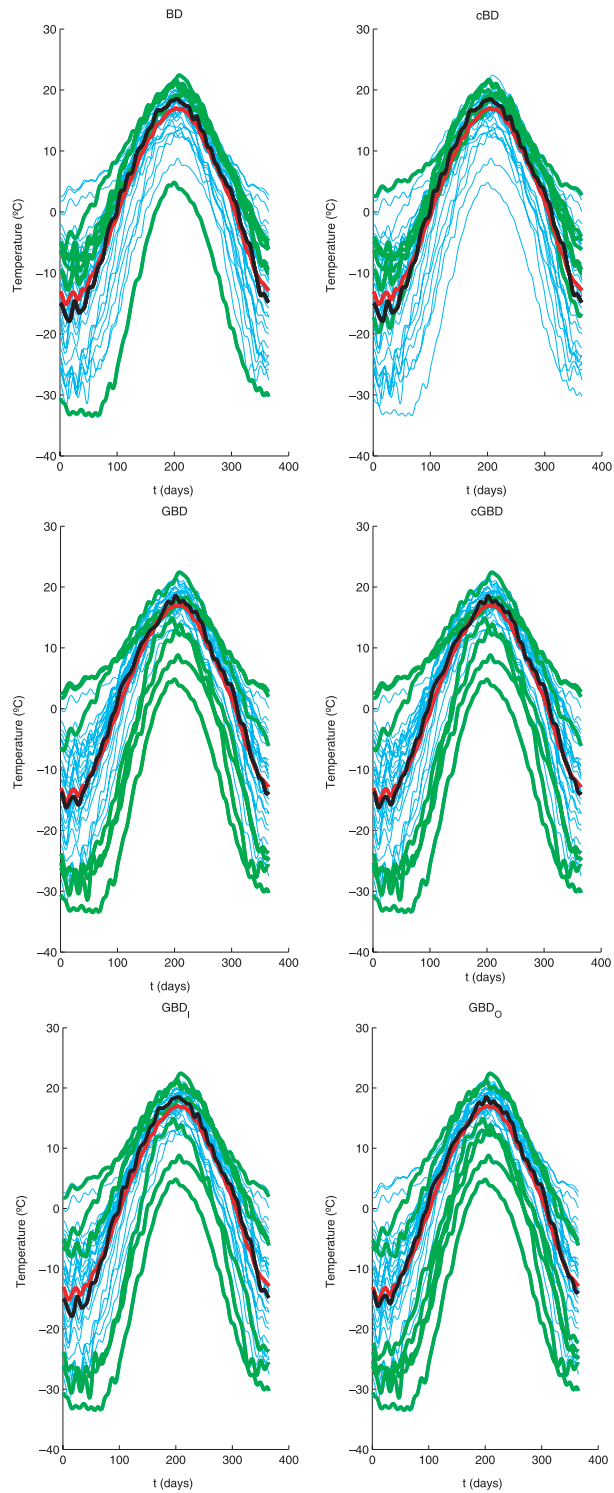

FIG 7. *Comparison of the 6 notions of depth. In each figure, we depict the mean curve (red), and median curve (black). The 20% least deep curves are plotted in green (wide lines), whereas the 80% most central curves are depicted in cyan (thin lines). Top panel: BD and cBD. Middle panel: GBD and cGBD. Bottom panel: $GBD_I$ and $GBD_O$.*

Our recommendation is that a set of data depths are used to screen the data for contaminations. A graphical examination of the data set can elucidate the potential outliers or extreme observations. One must make a case-by-case decision as to which functional shapes constitute outliers in a particular data set.

We propose a fast and simple re-sampling based data depth calculation procedure. The rank-orders induced by the re-sampling based method closely agree with the rank-orders induced by the full data. With the computationally efficient re-sampling based method, the new notions of depth can be used as building blocks in non-parametric functional analysis such as clustering and classification.

**Acknowledgments.** The authors would like to thank the reviewer for many insightful comments and helpful suggestions.

R. Jornsten is partially supported by NSF Grant DMS-0306360 and EPA Star Grant RD-83272101-0.

The work for this paper was conducted while S. López-Pintado was a postdoctoral researcher at Rutgers University, Department of Statistics.